\documentclass[prl,twocolumn,groupedaddress]{revtex4}
\usepackage[dvips]{color}
\usepackage{graphicx,color}
\usepackage{bm}        
\usepackage{amssymb}   
\usepackage{amsmath}
\usepackage{cases}
\usepackage{mathrsfs}

\begin{document}

\title{Possible Topological Superconducting Phases of MoS$_{2}$}

\author{Noah F. Q. Yuan $^{1}$, Kin Fai Mak$^{2,3}$}
\author{ K. T. Law$^1$} \thanks{phlaw@ust.hk}

\affiliation{$^1$Department of Physics, Hong Kong University of Science and Technology, Clear Water Bay, Hong Kong, China}
\affiliation{$^2$  Kavli Institute at Cornell for Nanoscale Science, Ithaca, NY 14853, USA}
\affiliation{$^2$  Laboratory of Atomic and Solid State Physics, Cornell University, Ithaca, NY 14853, USA}

\begin{abstract} Molybdenum disulphide (MoS$_2$)  has attracted much interest in recent years due to its potential applications in a new generation of electronic devices. Recently, it was shown that thin films of MoS$_2$ can become superconducting with a highest $T_{c}$ of 10K when the material is heavily gated to the conducting regime. In this work, using the group theoretical approach, we determine the possible pairing symmetries of heavily gated MoS$_2$. Depending on the electron-electron interactions and Rashba spin-orbit coupling, the material can support an exotic spin-singlet $p +ip $-wave-like, an exotic spin-triplet s-wave-like and a conventional spin-triplet $p$-wave pairing phases. Importantly, the exotic spin-singlet $p+ip$-wave phase is a topological superconducting phase which breaks time-reversal symmetry spontaneously and possesses non-zero Chern numbers where the Chern number determines the number of branches of chiral Majorana edge states. 
\end{abstract}


\maketitle

\emph{ \textbf{ Introduction}}--- A monolayer of Molybdenum disulfide (MoS$_2$) is a chemically stable 2D material similar to graphene. It consists of triangularly arranged Mo atoms sandwiched between two layers of triangularly arranged S atoms [\onlinecite{Bromley, Boker}]. Pristine monolayer MoS$_2$ is a direct band gap semi-conductor with a gap of about 1.8eV located at the two $K$ points of the Brillouin zone [\onlinecite{Mak1,Splendiani, Zhu, Shi, Zahid,Cappelluti}]. Due to its layered structure, chemical stability, relatively high mobility, strong spin-orbit coupling (SOC)  and the intrinsic massive Dirac gap, MoS$_2$ is considered a candidate material for next generation ultra-thin field effect transistors [\onlinecite{Kis, YZhang, QHWang, Bao, Mak2}] and valley-tronic devices [\onlinecite{Xiao, Mak3, Zeng}]. 

Interestingly, it was shown recently that thin films of MoS$_2$ can become superconducting when the material is heavily gated to the conducting regime where part of the conduction band near the $K$ points are occupied [\onlinecite{Ye, Taniguchi}]. At optimal gating, the superconducting transition temperature reaches 10K. Two possible pairing phases have been studied previously. They are the electron-phonon interaction induced $s$-wave pairing phase [\onlinecite{Ge, Rosner}] and the unconventional superconducting phase with opposite pairing signs for the electron near opposite $K$ points [\onlinecite{Roldan}]. As we show below, in the absence of Rashba SOC, these two superconducting phases are the only possible superconducting phases. However, due to the presence of the strong gating electric field in the experiment, which is of the order of $10$MeV/cm [\onlinecite{Ye}] and breaks the inversion symmetry, Rashba SOC can arise [\onlinecite{Ochoa, Jelena, Kormanyos}]. The Rashba SOC induces two superconducting phases which can be topologically non-trivial.

Particularly, using the group theoretical approach and solving the self-consistent gap equations, the possible superconducting pairing symmetries and the phase diagrams of  heavily gated MoS$_2$ thin films can be determined. We show that, in the presence of the Rashba SOC, an exotic \emph{spin-singlet} $p +ip$-wave-like pairing phase can be realized. This phase breaks time-reversal symmetry spontaneously and supports chiral Majorana edge states. Moreover, a more conventional spin triplet-singlet mixing $p \pm ip +s$-wave pairing phase can also be stablized by Rashba SOC. The properties of the topological superconducting phases are studied.

\begin{figure}
\centering
\includegraphics[width=3.2in]{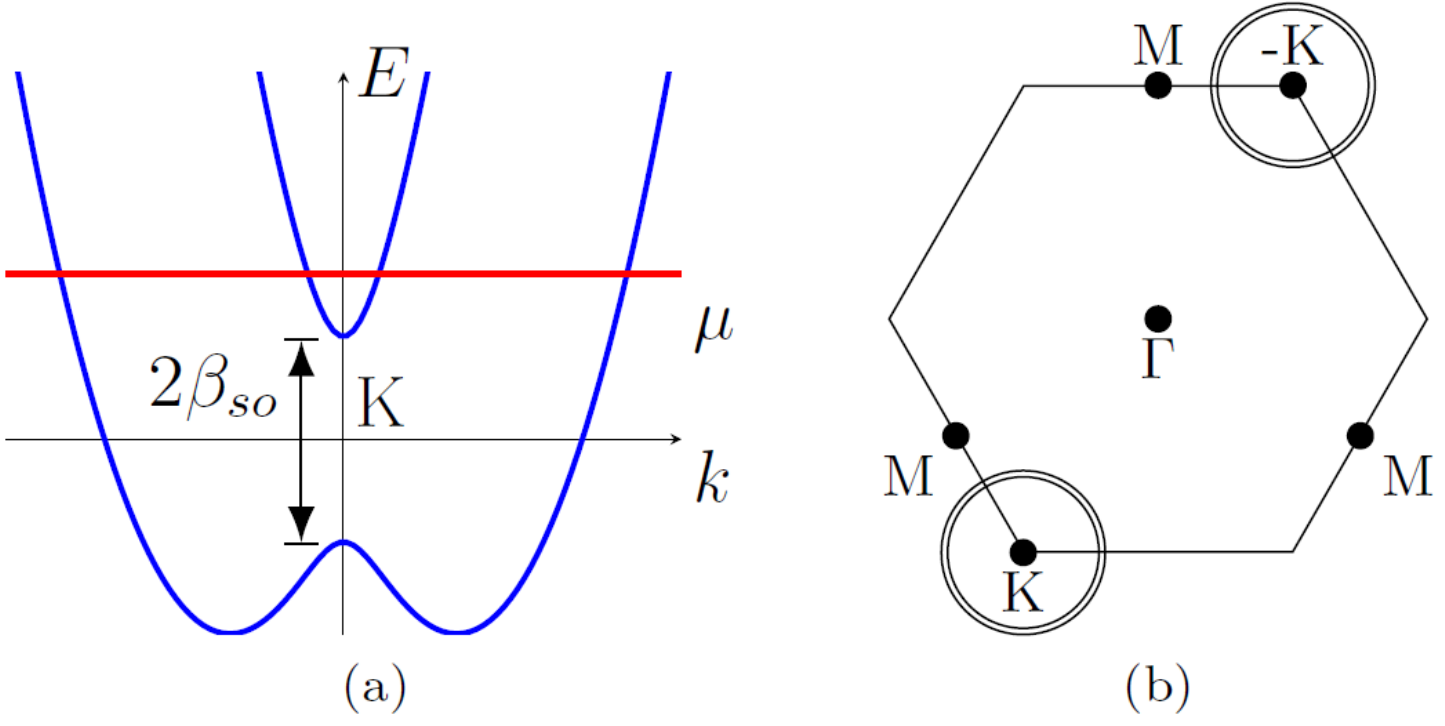}
\caption{a) The energy spectrum of a monolayer MoS$_2$ near the $K$ point. b) The Brillouin zone of MoS$_2$. We consider the regime in which the Fermi surfaces enclose the $\pm K$ points only. States near the time-reversal invariant $\Gamma$ and $M$ points are not occupied. }
\end{figure}

\emph{\textbf{Normal state Hamiltonian}}---
In recent experiments, thin films of MoS$_2$ become superconducting when the thin films are heavily gated such that part of the conduction band minimum near the $\pm K$ points are occupied [\onlinecite{Ye, Taniguchi}]. According to previous calculations [\onlinecite{Xiao, Mattheiss}], the conduction band minimum near the $\pm K$ points are dominated by the Mo 4$d_{z^2}$ orbitals. Therefore, we define the creation operators of 4$d_{z^2}$ electrons as $ c^{\dagger}_{s}$, where $s= \uparrow/\downarrow$ denotes spin. The monolayer MoS$_2$ respects the point group $C_{3v}$ symmetry.  The general Hamiltonian near the $\pm K$ points in the basis of $(c_{\bm k \uparrow}, c_{\bm k \downarrow})$ can be written as:
\begin{equation}  \label{Hamil}
H_{0}(\bm k + \epsilon \bm K)= \frac{{|\bm k|}^{2}}{2m} -\mu + \alpha_{R} \bm g (\bm k) \cdot \bm \sigma +
\epsilon \beta_{so} \sigma_z .    
\end{equation}
Here, $\epsilon = \pm$ is the valley index, $m$ is the effective mass of the electrons, $\mu$ is the chemical potential measured from the conduction band bottom when SOC is omitted.  The Rashba SOC strength due to the gating electric field, which breaks the mirror symmetry in the $z$ direction, is denoted by $\alpha_{R}$ and the Rashba vector is $\bm g (\bm k)= (k_y, -k_x, 0) $.  The intrinsic SOC strength due to the coupling between the lowest conduction band and the other bands is $\beta_{so}$ [\onlinecite{Xiao, Kormanyos2}]. The $\beta_{so}$ term plays the role of an effective Zeeman field and splits the spin up and spin down bands at the $\pm K$ points. However, it is important to note that this intrinsic SOC strength has opposite signs at the $K$ and $-K$ points such that the total Hamiltonian respects time-reversal symmetry. A schematic picture of the band structure near the $K$ point is depicted in Fig.1a. The Fermi surfaces near the $K$ points are depicted in Fig.1b. We show below that both of the SOC terms $\alpha_{R}$ and $\beta_{so}$ are important for determining the topological properties of the superconducting phases.


\emph{\textbf{Possible superconducting pairing phases}}--- To study the possible superconducting phases of the system, we denote the interacting Hamiltonian of the system as 
\begin{equation}
\begin{array}{ll}
H_{\text{int}}=\frac{1}{2}\sum V_{s_{1}s_{2}s_{3}s_{4}}(\bm k,\bm k')
c_{\bm k s_{1}}^{\dagger} c_{-\bm k s_{2}}^{\dagger} c_{\bm k' s_{3}} c_{-\bm k' s_{4}},
\end{array}
\end{equation}
where $V$ parametrizes the interaction strength of the electrons. Due to the $C_{3v}$ point group symmetry of the monolayer MoS$_2$, the mean field superconducting pairing matrix resulting from $H_{\text{int}}$ can be classified according to the irreducible representations of $C_{3v}$. Denoting $A_1$ as the trivial representation and $E$ as the two-dimensional representation of $C_{3v}$, respectively, we found that the corresponding superconducting pairing matrix $\Delta_{\Gamma}$ in the $A_1$ and the $E$ irreducible representations can be written as:
\begin{eqnarray}   \label{gap}
&&\Delta_{\Gamma}(\bm k)=     
\\
&&\left\{
\begin{array}{ll}
[s_{A_1,1}\psi_{A_1}(0)+s_{A_{1},2}\psi_{A_{1}}(\bm k) \\ +t_{A_{1},z}\bm d_{A_{1},z} \cdot\bm\sigma
+t_{A_{1},xy}\bm d_{A_{1},xy} \cdot\bm\sigma ]i\sigma_{y}&\Gamma =A_{1}
\\
\\
\sum_{m= \pm} [s_{E,m}\psi_{E,m}+t_{E,m}\bm d_{E,m}\cdot\bm\sigma ]i\sigma_{y}&\Gamma =E .
\end{array}
\right.
\nonumber
\end{eqnarray}

\begin{table}
\centering
\begin{tabular}{l*{3}{c}r} \hline
$\Gamma$      & singlet         & triplet \\
\hline\hline
$A_{1}$                    & $\psi_{A_1}(0) = C (\bm k =0) $       &  $\bm d_{A_1, z} =S(\bm k)\bm z$ \\
                          & $\psi_{A_1}(\bm k) = C(\bm k)$ & $\bm d_{A_1, xy} =S_{-}(\bm k)\bm x_{+}-S_{+}(\bm k)\bm x_{-}$ \\
    
\hline
$E$ &$\psi_{E,+} =C_{+}(\bm k)$&$\bm d_{E,+}=S_{+}(\bm k)\bm z$\\
     &$\psi_{E, -} =C_{-}(\bm k)$&$\bm d_{E,-}=S_{-}(\bm k)\bm z$\\
      \hline
\end{tabular} 
\caption{Possible pairing phases of monolayer MoS$_2$. They are labeled by the irreducible representations of $C_{3v}$. Here, a set of basis functions are introduced: 
$
C(\bm k)=\sum_{j=1}^{3}\cos(\bm k\cdot\bm R_{j}),
C_{+}(\bm k)=\sum_{j=1}^{3}\omega^{j-1}\cos(\bm k\cdot\bm R_{j}),
S(\bm k)=\sum_{j=1}^{3}\sin(\bm k\cdot\bm R_{j}),
S_{+}(\bm k)=\sum_{j=1}^{3}\omega^{j-1}\sin(\bm k\cdot\bm R_{j})
$
and $C_{-}(\bm k)=C_{+}^{*}(\bm k),S_{-}(\bm k)=S_{+}^{*}(\bm k)$. The phase factor is $\omega =\exp(2\pi i/3)$. The bonding vectors of Mo atoms are $\bm R_{1}=a\bm x, \bm R_{2}=a(-\bm x/2+\sqrt{3}\bm y/2), \bm R_{3}=a(-\bm x/2-\sqrt{3}\bm y/2)$ with lattice constant $a=1$. Orthonormal complex vectors $\bm x_{+}=(\bm x+i\bm y)/\sqrt{2}$ and $\bm x_{-}=(\bm x-i\bm y)/\sqrt{2}$ are introduced.}
\label{Pairings}
\end{table}

Here, $s_{\Gamma}$ and $t_{\Gamma}$ are constants which denote the spin-singlet and spin-triplet pairing strengths respectively.  The basis functions $\psi_{\Gamma}$ and $ {\bm d_{\Gamma}} $ of the irreducible representations of the point group $C_{3v}$ are shown in Table I.  

In the $A_{1}$ representation, $\psi_{A_1}(0)$ represents the conventional $\bm k$-independent $s$-wave pairing. Near the $K$ points, $\psi_{A_1}(\bm K + \bm k) \approx  -\frac{3}{2}+\frac{3}{8}|\bm k|^2 $ denotes the extended $s$-wave pairing. On the other hand, the $\bm d$-vector $\bm d_{{A_1},z}(\bm k + \epsilon \bm K) \approx -  \epsilon \sqrt {3}\frac{3}{2} \bm z $ parametrizes a  spin-triplet pairing phase as the condition $\bm d_{{A_1},z}(\bm k)= - \bm d_{{A_1},z}(-\bm k)$ is satisfied. Interestingly, the pairing amplitudes are approximately $\bm k$ independent near  $\pm K$ points but with opposite sign. 

This exotic  \emph{spin-triplet} $s$-wave-like pairing phase is possible due to the triangular lattice structure of the Mo atoms. It appears when the nearest neighbour electrons, at sites $i$ and $j$ respectively, satisfy the pairing relation $\langle c_{i, \uparrow} c_{j, \downarrow} + c_{i, \downarrow} c_{j, \uparrow} \rangle = \Delta_0 e^{i \alpha_{ij}}$, where $\Delta_0$ is a constant and the phase factors $\alpha_{ij}$ are shown in Fig.2a. The possibility of this pairing phase is pointed out in Ref.[\onlinecite{Roldan}].

\begin{figure}\label{phase}
\centering
\includegraphics[width=3.2in]{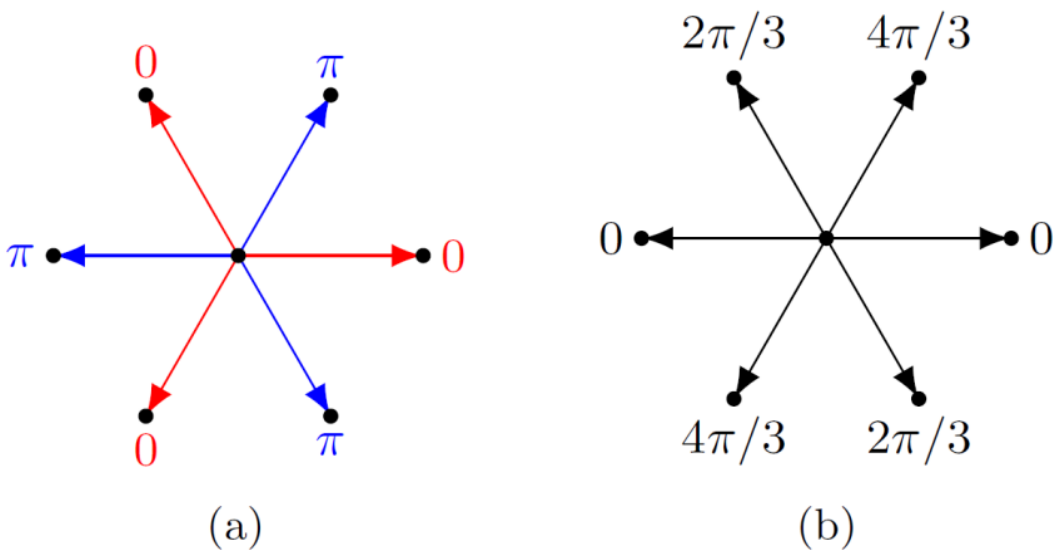}
\caption{a) The nearest neighbour pairing phases $\alpha_{ij}$ in the spin-triplet, $s$-wave-like, $\bm d_{A_1,z}$-phase are shown. The black dots represent the lattice sites with the central dot represents site $i$. b) The nearest neighbour pairing phases $\beta_{ij}$ of the  spin-singlet, $p$-wave-like,$\psi_{E,+}(k)$-phase. }
\end{figure}

On the other hand, near the $\pm K$ points, $\bm d_{A_1, xy}(\bm k + \epsilon \bm K) \cdot \bm \sigma \propto \sigma_x k_y -\sigma_y k_x $ and it parametrizes a spin-triplet $p \pm ip$-pairing phase, similar to the non-centrosymmetric superconductor cases [\onlinecite{Sigrist,Yip}], $\bm d_{A_1, xy} $ is parallel to the Rashba vector $\bm g (\bm k)$ in Eq.1 and this superconducting phase can be stabilized by Rashba SOC.

For the two-dimensional $E$ representation, the spin-singlet basis functions $\psi_{E, \pm} (\bm k+ \epsilon \bm K) \approx \epsilon 4\sqrt{3} (k_x \pm ik_y)$ near $\pm K$ points. It is important to note that $\psi_{E,\pm}(\bm k) = \psi_{E,\pm}(- \bm k)$ are even in $k$ such that Fermi statistics is satisfied. However, near  $\pm K$ points, the pairing has $p+ip$-wave characteristics. The $\psi_{E,+}$ pairing is realized when the nearest neighbour singlet pairing amplitudes satisfy the relation $\langle c_{i, \downarrow} c_{j, \uparrow} - c_{i, \uparrow} c_{j, \downarrow}  \rangle = \Delta_{0} e^{i \beta_{ij}} $, where the phase $\beta_{ij}$ is depicted in Fig.1b. The $\psi_{E,-}$ pairing is realized when $\beta_{ij} \to -\beta_{ij}$. We show in later sections that this \emph{spin-singlet} $p+ip$-wave phase is a topological superconducting phase. 

The remaining pairing phase characterized by  $\bm d_{E, \pm} (\bm k + \epsilon\bm K) \approx -\frac{3}{4}(k_x \pm i k_y) \bm z $ is a spin-triplet $p$-wave pairing phase. This is a topological phase similar to the $^3$He A-phase. Unfortunately, as we discuss below, this pairing is not energetically favourable in MoS$_2$.

\emph{\textbf{Phase diagrams}}--- To determine the stability of different superconducting phases and the singlet and triplet pairing amplitudes, we solve the linearized gap equations 
\begin{equation}
\begin{array}{ll}
\Delta_{ss'}(\bm k)&=k_{B}T_{c}\sum_{ns_{1}s_{2}\bm k'}V_{s'ss_{1}s_{2}}(\bm k,\bm k')\times\\
&[G_e (\bm k',i\omega_{n})\Delta (\bm k')G_h (\bm k',i\omega_{n})]_{s_{1}s_{2}},
\end{array}
\end{equation}
where $T_c$ is the superconducting transition temperature.  The normal state Matsubara Green's functions for electrons and holes are denoted as $ G_{e}(\bm k,i\omega_{n})=[i\omega_{n}-H_{0}(\bm k)]^{-1} $ and $G_{h}(\bm k,i\omega_{n})=[i\omega_{n}+H^{*}_{0}(-\bm k)]^{-1}$, respectively. By expanding the interaction terms using the basis functions [\onlinecite{Sigrist2}], 
\begin{equation}
\begin{array}{ll}
V_{s'ss_{1}s_{2}}(\bm k,\bm k')=-v_{0}/A(i\sigma_{y})_{ss'}(i\sigma_y )_{s_{1}s_{2}}\\
-v_{1}/A\sum_{\Gamma ,m}\lbrace \psi_{\Gamma ,m}(\bm k)\psi_{\Gamma ,m}(\bm k')(i\sigma_{y})_{ss'}(i\sigma_y )_{s_{1}s_{2}}\\
+[\bm d_{\Gamma ,m}(\bm k)\cdot\bm\sigma i\sigma_y]_{ss'}[\bm d_{\Gamma ,m}(\bm k')\cdot\bm\sigma i\sigma_y]_{s_{1}s_{2}}\rbrace ,
\end{array}
\end{equation} 
with the sample area $A$, we can determine the superconducting phase with the highest superconducting transition temperature $T_c$ as a function of $v_0$ and $v_{1}$, where $v_0$ and $v_1$ denote the on-site and nearest neighbour interaction strengths of the electrons respectively. Positive (negative) values of $v_{i}$ represent attractive (repulsive) interactions and they are determined by electron-phonon or other electron-electron interactions. We choose $m=0.5 m_e$, $\beta_{so}= 2$meV [\onlinecite{Kormanyos2}], in the normal state Hamiltonian such that the energy spectrum near the $K$ points matches the DFT results [\onlinecite{Kormanyos}], where $m_e$ is the electron mass. The resulting phase diagrams with and without Rashba SOC are presented in Figs.3a and 3b respectively.

\begin{figure}
\centering
\includegraphics[width=3.2in]{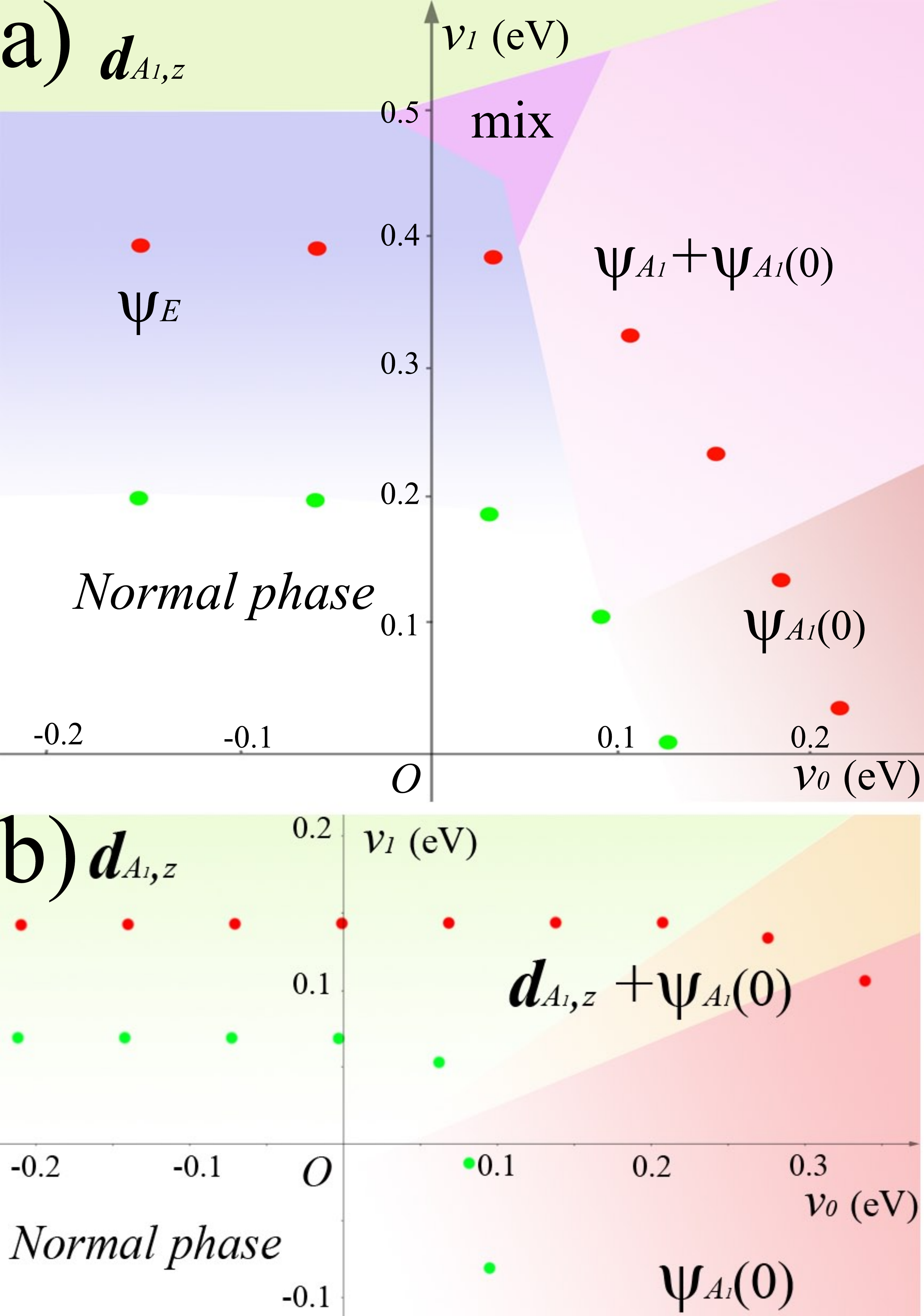}
\caption{\label{pd} The superconducting phase diagrams of MoS$_2$. The phases are labeled by the basis functions in Table I characterizing the phases. a) With finite Rashba SOC $m\alpha_R^{2}/2 =8$meV. The mix phase is a mixture of $\psi_{A_1}(0),\psi_{A_1}(\bm k),\bm d_{A_1,xy}$ and $\bm d_{A_1,z}$ phases. b) $\alpha_R =0$. The red (green) dots denote the regime with critical temperature of $10$K ($4\times 10^{-3}$K). }
\end{figure}

It is evident from Fig.3a that, in the presence of Rashba SOC, both the $E$ and the $A_1$ phases are possible depending on the interaction strengths $v_0$ and $v_1$. By solving Eq.4, we note that the $E$ phase, which is labeled by $\psi_{E}$ in Fig.3a, is favourable when the on-site attraction is weak or repulsive and nearest neighbour interaction is relatively strong. We also note that $t_E$, which gives the triplet pairing strength, is negligible in the $E$ phase. Therefore, the singlet pairing component dominates and the pairing matrix can be written as $\Delta_{E} = \sum_{\pm} s_{E,\pm}\psi_{E, \pm}  i \sigma_y$. Moreover, through the fourth order Gingzburg-Landau analysis [\onlinecite{Sigrist3}], we found that the free energy is minimal only when $s_{E,+}$ or $s_{E,-}$ is zero. As a result, time-reversal symmetry is spontaneously broken in the $E$ phase. We show in the next section that this phase is characterized by Chern numbers and supports Majorana edge states.

As expected, strong on-site attraction $v_0$ favours the usual $\bm k$-independent $s$-wave pairing, denoted by $\psi_{A_1}(0)$ in Fig.3a. When $v_0$ and $v_1$ are both attractive and comparable in magnitude, both the extended s-wave pairing $\psi_{A_1}(\bm k)$ and $\psi_{A_1}(0)$ are significant. It is interesting to note that there is a regime in which all the spin-singlet pairings and the spin triplet $\bm d_{A_1,xy}$ and $\bm d_{A_1,z}$ pairings in the $A_1$ representations are mixed. This is possible as the Rashba SOC breaks the inversion symmetry. When $v_{1}$ is strongly attractive, the spin-triplet $\bm d_{A_1,z}$-pairing dominate.

Importantly, MoS$_2$ can be chemically doped to the superconducting regime [\onlinecite{Ye}]. In this case, the mirror symmetry in the $z$-direction can be respected which results in zero Rashba SOC. The phase diagram in the absence of the Rashba SOC term is shown in Fig.3b. It is evident that only the pairing phases belonging to the $A_1$ representation appear in this case. When on-site attraction is strong, the conventional $s$-wave component is favoured. When the nearest-neighbour attraction is strong, the exotic spin-triplet $s$-wave-like pairing characterised by $\bm d_{A_1,z}$ becomes dominant. Both of these phases have been studied previously in which Rashba SOC is ignored. It is interesting that even when Rashba SOC is absent, the spin-singlet and spin-triplet mixing $\bm d_{A_1,z} + \psi_{A_1} (0)$-phase can be realized in a narrow parameter regime since the in-plane mirror symmetries are broken by the $\beta_{so}$ term.

\emph{\textbf{Topological phases}}--- In the previous sections, we pointed out that interesting phases, such as the exotic spin-singlet $p$-wave characterized by $\psi_{E}(\bm k)$ and the spin-triplet $s$-wave characterized by $\bm d_{A_1, z}$ can possibly be realized in the MoS$_2$. In this section, we study the topological properties of these pairing phases.

In general, the BdG Hamiltonian can be written as 
\begin{equation}
\begin{array}{l}
H_{\text{BdG}}(\bm k)=\left(
\begin{array}{cc}
 H_{0}(\bm k) & \Delta_{\Gamma}(\bm k) \\
\Delta^{\dagger}_{\Gamma}(\bm k) & -H_{0}^{*}(-\bm k)
\end{array} \right).
\end{array}   \label{Hk}
\end{equation}

In the $E$ phase with $ \Delta_{\Gamma} =  s_{E,+} \psi_{E,+} i \sigma_{y}$, $H_{\text{BdG}}$ breaks time-reversal symmetry and the Hamiltonian can be classified by Chern numbers [\onlinecite{Schnyder, Teo}]. The Chern number can be written as
\begin{equation}
N_{\text{Chern}}= \frac{1}{2\pi} \sum_{E_{n}<0} \int d^{2} \bm k \partial_{x} a^{n}_{y} - \partial_{y} a^{n}_{x},
\end{equation}
where $a^{n}_{i} = -i \langle n, \bm k| \partial_{k_{i}} | n, \bm k \rangle $, $n$ is the band index, $| n, \bm k \rangle $ denotes an eigenstate of the $H_{\text{BdG}}(\bm k)$, and the summation is over all the bands with negative energy $E_n$. It can be shown that:
\begin{numcases}
{N_{\text{Chern}}=} \nonumber
2&$ |\mu| < |\beta_{so}|$ \\   \nonumber
4&$ \mu > |\beta_{so}| $ \\
0 & otherwise.
\end{numcases}

\begin{figure}
\centering
\includegraphics[width=3.2in]{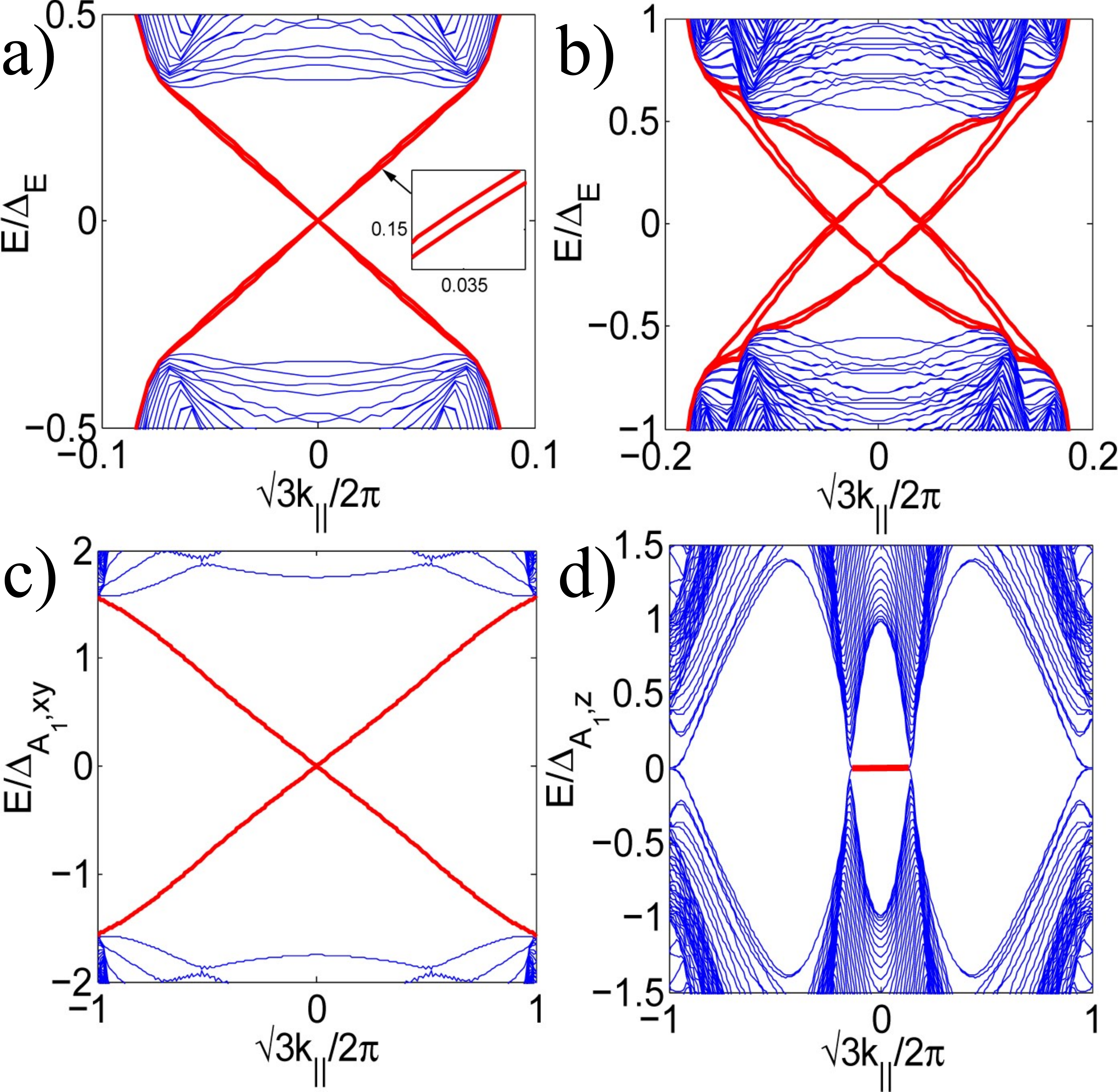}
\caption{\label{spec} The spectra of superconducting MoS$_2$ with open zig-zag edges. The parameters are $\beta_{so} =1$, $\Delta =1$ in all figures. a) and b): The spectra in the $E$ phase with $ m=1/30$, $\alpha_R =1$.  In a), $\mu =0$ such that $N_{\text{Chern}}=2$. In b),  $\mu =4$ such that $N_{\text{Chern}}=4$. The edge states are highlighted in red. c) $ m=1/15$, $\mu = 48$ in the $\bm d_{A_1,xy}$-phase, which can support Majorana edge states. d)  $\alpha_R =0$, $ m=1/15$, $\mu = 11$ in the $\bm d_{A_1,z}$ -phase, which can be a nodal topological phase with Majorana flat bands.}
\end{figure}

To verify the topological nature of the superconducting states, we construct a tight-binding model for an infinitely long strip of MoS$_2$ with finite width. Since the Mo layer has a triangular lattice, both flat and zig-zag edges are allowed. The energy spectrum of the tight-binding model as a function $k_{\parallel}$ is shown in Fig.4, where $k_{\parallel}$ is the momentum quantum number parallel to the zig-zag edge. It is evident that when the $N_{\text{Chern}}=2$ (Fig.4a), there are two chiral edge states propagating on each edge of the system. When the chemical potential is increased such that $N_{\text{Chern}}=4$ (Fig.4b), there are four chiral edge states propagating on each edge. Similar edge states for flat edges can also be found. Since this phase breaks time-reversal symmetry, we expect that it can be probed by $\mu$SR measurements [\onlinecite{Luke, Biswas}]. 

It is interesting to note that superconducting topological states with chiral Majorana edge modes have been proposed to exist in graphene with $d+id$-wave pairing [\onlinecite{Annica1, Daoxin, Levitov, Annica2, Annica3}], even though intrinsic superconductivity in graphene is yet to be found. Applying a Zeeman field can change the Chern number in $d+id$ pairing graphene [\onlinecite{Annica3}] while the $\beta_{so}$ term in MoS2 plays the role of an effective magnetic field at each K point which can change the Chern numbers of the system.

Another interesting phase of MoS$_2$ is the time-reversal invariant spin-triplet $p$-wave pairing phase characterized by $ \Delta_{\Gamma} = t_{A_1,xy} \bm d_{A_1, xy} \cdot \bm \sigma i\sigma_{y}$. This phase is not prominent in Fig.3a. However, since $ \bm d_{A_1, xy} \parallel \bm g(\bm k)$, this pairing becomes dominant when Rashba SOC is strong, similar to the the case in certain non-centrosymmetric superconductors [\onlinecite{Sigrist,Yip}]. Unfortunately, in the superconducting regime studied experimentally, the Fermi surface is expected to enclose the $\pm K$ points only, which have energy far below the time-reversal invariant $\Gamma$ and the $M$ points depicted in Fig.1b. Therefore, the system cannot be a topologically non-trivial time-reversal invariant topological superconductor in which the surfaces have to enclose the time-reversal invariant points [\onlinecite{Qi}]. This topological phase can be relevant to other superconducting transition metal dichacogenides in which the $M$ points have similar energy to the $K$ points.  As depicted in Fig.4c, Majorana edge states appear when the chemical potential is above the energy level of the $M$ points, even though we expect that this regime not to be realistic in MoS$_2$.

It is worth mentioning that the exotic spin-triplet $s$-wave-like pairing phase characterized by $ \Delta_{\Gamma} = t_{A_1,z}\bm d_{A_1, z}\cdot \bm \sigma i\sigma_{y}$ is a nodal topological phase which supports Majorana flat bands [\onlinecite{Tanaka, Sato, Schnyder2, Fa, Wong}], given that the chemical potential is above the energy level at the $M$ point. The energy spectrum of an infinite strip of MoS$_{2}$ with zig-zag edge in the nodal topological regime is shown in Fig.4d. It is evident that Majorana flat bands appear in this regime. This nodal topological phase can be relevant to other transition metal dichacogenides.

\emph{\textbf{Conclusion}}--- In this work, we show that Rashba SOC induces new superconducting phases beyond the conventional $s$-wave pairing and the spin-triplet $s$-wave-like pairing studied previously [\onlinecite{Roldan}]. Particularly, the spin-singlet $p$-wave-like phase, which breaks time-reversal symmetry, is a topological superconducting phase. We also found a spin-triplet $p$-wave phase which can be stabilized by Rashba SOC and this phase is expected to be in the topologically trivial regime in MoS$_2$. The topological properties of the unconventional phases are studied and the exotic superconducting phases found in this work can also be relevant to other transition metal dichacogenides.

\section{Acknowledgement}
We thank Jeffrey Teo, Hung Yao and Jian-Ting Ye for illuminating discussions. KTL and NFQY thank the support from HKRGC through Grants 605512 and 602813 and CRF3/HKUST/13G. 

\end{document}